\documentclass[11pt]{article}
\usepackage{graphicx}

\DeclareMathSizes{10}{8.5}{7}{7}

\newcommand{\be}{\begin{equation}}
\newcommand{\bea}{\begin{eqnarray}}
\newcommand{\ee}{\end{equation}}
\newcommand{\eea}{\end{eqnarray}}

\def\K{\mbox{\bf K}}
\def\Z{\mbox{\bf Z}}

\def\x{\mbox{{\bf x}}}

\def\k{\mbox{{\bf k}}}
\def\g{\mbox{{\bf g}}}

\def\Y{\mbox{$\bf{Y}$}}
\def\Z{\mbox{$\bf{Z}$}}
\def\Yhat{\mbox{$\bf{\hat{Y}}$}}

\def\y{\mbox{$\bf{y}$}}
\def\v{\mbox{$\bf{v}$}}
\def\V{\mbox{$\bf{V}$}}

\def\w{\mbox{$\bf{w}$}}

\def\z{\mbox{$\bf{z}$}}

\def\p{\mbox{$\bf{p}$}}

\def\Phib{ \mbox{\boldmath{$\Phi$}} }
\def\phib{ \mbox{\boldmath{$\phi$}} }

\def\Psib{ \mbox{\boldmath{$\Psi$}} }
\def\psib{ \mbox{\boldmath{$\psi$}} }

\def\f{\mbox{$\bf{f}$}}

\def\K{\mbox{$\bf{K}$}}

\begin{document}

\begin{center}

{\Large Effective Actions for Ensemble Data Assimilation}\\
\bigskip
\bigskip
Henry D. I. Abarbanel\\
\bigskip
Department of Physics\\ and\\ Marine Physical Laboratory (Scripps Institution of Oceanography),\\ University of California,
San Diego,\\ 9500 Gilman Drive, Mailcode 0402,\\ La Jolla, CA
92093-0402   USA\\
habarbanel@ucsd.edu
\end{center}

\date{\today}

\begin{abstract}
Ensemble data assimilation is a problem in determining the most likely phase space trajectory of a model of an observed dynamical system as it receives inputs from measurements passing information to the model. Using methods developed in statistical physics, we present effective actions and equations of motion for the mean orbits associated with the temporal development of a dynamical model when it has errors, there is uncertainty in its initial state, and it receives information from measurements. If there are correlations among errors in the measurements they are naturally included in this approach.
\end{abstract}

Assimilating the information in observed data into models of a dynamical system when there are errors in the measurements, errors in the models, and uncertainty about the precise state of a system when the assimilation process begins has stimulated discussions about placing data assimilation in a probabilistic setting~\cite{lorenc86,pham}.  The goal is to calculate the probability of a model state vector $\y(t_m)$ to be at a certain location in state space at time $t_m$ conditioned on the observation of data at 
$t_0, t_1, ..., t_m$. A recursive formula for this conditional probability has been discussed. We give an information theoretic interpretation of this recursion relation and indicate how it can be systematically approximated within a framework that is suggested by developments in statistical physics. While approximations can be made at almost any stage, we show how to systematically represent the effective action associated with the procedure. We also indicate how one can estimate unknown model parameters.

This subject is broadly addressed in the sciences. We have examined specific research seeking to estimate parameters and states in neurobiology~\cite{huys}, systems biology~\cite{ash}, atmospheric and oceanic 
sciences~\cite{evensen,kalnay,lorenc}, biomedical 
engineering~\cite{horv}, cell biology~\cite{beard}, chemical engineering~\cite{xiong}, toxicology~\cite{schr}, coastal and estuarine modeling~\cite{yang}, wastewater treatment~\cite{mueller}, biochemistry~\cite{dochain}, 
and immunology~\cite{swam} as examples. The issue of constructing observers in control theory~\cite{nij1} also deals with state estimation from observed data. The literature has focused on disciplinary details of models, metrics for errors of model output, model errors, measurement errors, and numerical methods.

We review the formulation of data assimilation in an ensemble or probabilistic sense to establish notation and to provide a framework for our subsequent discussion.
We begin with an observed dynamical system with state variables $\x(t_n) = \x(n)$. Over an observation or assimilation window at each of the discrete times $t_n: \{t_0,t_1,...,t_{m}\}$ L functions of the state variable $z_1(t_n) = h_1(\x(t_n)), ..., z_L(t_n) = h_L(\x(t_n))$ are reported. This provides us with $m + 1$ observed L-dimensional vectors $\z(t_n) = \z(n)$. 
From physical arguments we construct a D-dimensional model of this dynamical system with state variables 
$\w(t_n) = \w(n)$ and with a dynamical rule $\w \to \g(\w,\p)$ taking the state at time $t_n$ to the state at time 
$t_{n+1}:\;\w(n+1) = \g(\w(n),\p)$. $\p = \{p_1,p_2,...\}$ are fixed parameters of the model. D is typically much greater than 
L.  

We wish to choose the model, the initial conditions $\w(0)$, and the parameters $\p$ so that at the measurement times the model state is such that $h_l(\w(n)) = z_l(n);\;l=1,2,...,L$. Because D $\gg$ L, we must estimate the unobserved state variables as well as any unknown model parameters in order to predict forward from the end of the data window $t > t_m$.

Now change variables from $\w$-space to $y_l(n) = h_l(\w(n));\;l=1,2,...,L$ and $y_a(n) = w_a(n);\;a = L+1,L+2, ...,D$. The dynamics in $\y$ space is 
$\y(n+1) = \f(\y(n),\p).$
Fixed parameters in the functions $y_l = h_l(\w)$ are now included in the dynamical map $\y \to \f(\y,\p)$, and any estimation procedure may be asked to determine them as well.

There are errors in the observations. At each time $t_0, t_1, ..., t_m$ there is a distribution of possible observations from which the measured values are drawn. Since there are many realizations of $z_l(t_0)$, there are many possible trajectories for the model to track. 

We want to evaluate the conditional probability that at time $t_m$ the state of the model is $\y(m)$ given the specific sequence of observations 
$\z(0),\z(1), ..., \z(m)$. $\z(n)=[z_1(n),...,z_L(n)]$. The conditional probability is denoted
$P(\y(m)|\z(m),\z(m-1),...,\z(0))$. Its dependence on $\p$ and $\y(0)$ is not shown.
Introducing $\Z(m) = \{\z(m),\z(m-1),...,\z(0)\}$, write $P(\y(m)|\z(m),\z(m-1),...,\z(0)) = P(\y(m)|\Z(m))$.

By the definition of conditional probabilities we have the identity
\be 
P(\y(m)|\Z(m)) = 
\frac{P(\z(m)|\y(m),\Z(m-1))P(\y(m)|\Z(m-1))}{P(\z(m)|\Z(m-1))}
\ee
which is a known, useful result~\cite{lorenc,hamill,hansen}. We reproduce it to introduce our notation and to provide our starting point. 

It is informative to rewrite this identity as
\bea 
&&P(\y(m)|\Z(m)) =   \nonumber \\
&&\biggl \{ \frac{P(\y(m),\z(m)|\Z(m-1))}{P(\z(m)|\Z(m-1))P(\y(m)|\Z(m-1))} \biggr \} \nonumber \\
&&P(\y(m)|\Z(m-1)). 
\eea 
The coefficient in curly brackets is the exponential of the conditional mutual information between 
the L-dimensional measurement $\z(m)$ and the D-dimensional model output $\y(m)$, conditioned on the previous observations $\Z(m-1)$~\cite{fano}. 
We call it 
\be MI(\y(m),\z(m)|\Z(m-1)) = 
\log [ \frac{P(\y(m),\z(m)|\Z(m-1))}{P(\z(m)|\Z(m-1))P(\y(m)|\Z(m-1))} ].
\ee
This answers the question: how much (in bits using $\log_2$) do we learn about $\y(m)$ from observing $\z(m)$, given we have already observed $\Z(m-1)$. 
$MI(\y(0),\z(0)|\Z(-1))] = \log  [ \frac{P(\y(0),\z(0))}{P(\y(0))\,P(\z(0))} ]$.

One often approximates the conditional mutual information assuming that errors in measurements are not correlated at different observation times. This makes $MI(\y(m),\z(m)|\Z(m-1))$ independent of $\Z(m-1)$, but the assumption is not needed for the general discussion. Including such correlations may be physically important~\cite{hamill}.

We have assumed that the description of the dynamics by the state vector $\y(m)$ represents the full set of dynamical variables. Thus the process $\y(m) \to \y(m+1)$ is Markov: $\y(m+1)$ depends only on $\y(m)$. 
We may use the Chapman-Kolmogorov equation~\cite{papou} to write
$
P(\y(m)|\Z(m-1)) = 
\int d^Dy(m-1) P(\y(m)|\y(m-1)) P(\y(m-1)|\Z(m-1)). 
$
For deterministic dynamics the transition matrix $P(\y(n+1)|\y(n)) = \delta^D(\y(n+1) - \f(\y(n),\p))$.

Combining these statements, we write a recursion relation for moving forward in time 
$
P(\y(m)|\Z(m)) = \exp[MI(\y(m),\z(m)|\Z(m-1))] 
\int d^Dy(m-1) 
P(\y(m)|\y(m-1)) P(\y(m-1)|\Z(m-1)).
$

Iterating back to $t_0$ we have 
\bea
&&P(\y(m)|\Z(m)) = \nonumber \\
&&\int \prod_{n=0}^{m-1}d^Dy(n) 
\exp[TMI(\Y,\Z(m))] \nonumber \\
&&P(\y(n+1)|\y(n)) P(\y(0)),
\eea
with $\Y = \{\y(m), \y(m-1), ..., \y(0)\}$ and $TMI(\Y,\Z(m)) = \sum_{n=0}^m MI(\y(n),\z(n))|\Z(n-1)).$

$TMI$ is the sum of the conditional mutual information associated with the observations $\z(n)$ and the model 
states $\y(n)$, conditioned on the observations previous to that time location. 
It represents the total information passed to the sequence of states 
$\Y$ by the measurements at $\Z(m)$. 

This is our basic formula for assimilating data measured at $\{t_0, t_1, ..., t_m\}$ into guiding a model trajectory $\y(n)$
and providing that model with information on any unknown parameters in the model and on the values of the unobserved state variables of the model.

Introduce the cumulant generating function
\bea
&&\exp[C_m(\K)] = \\
&& \int \prod_{n=0}^m d^Dy(n) \exp[ \sum _{n=0}^m \k(n) \cdot \y(n)] \exp[-A_0(\Y,\Z(m))], \nonumber
\eea
where
\bea
&&-A_0(\Y,\Z(m)) = TMI(\Y,\Z(m))  \\
&&+ \sum_{n=0}^{m-1} \log[P(\y(n+1)|\y(n))]+ \log[P(\y(0))], \nonumber
\eea
and this is an `action' for motion along the orbit $\Y$. We have associated a $\k(n)$ with each location in time where a measurement is made. The collection of these is denoted by $\K = \{\k(m), \k(m-1),..., \k(0)\}$. In statistical physics and quantum theory these are currents that produce excitations from the ground state. Here they are useful tools for analyzing the system at hand.

The conditional mean orbit is found as
$
<\y(n)>\, = \frac{\partial C_m(\K)}{\partial \k(n)}|_{\K = 0},
$
and the moments about this mean are determined by higher derivatives of $C_m(\K)$ at $\K = 0$..

It is very useful to define the $\K$ dependent trajectory $\phib(n)$ via $\phib(n) = \frac{\partial C_m(\K)}{\partial \k(n)}$
and with this to make the transformation of variables from the multipliers $\K$ to $\Phib = \{\phib(m),\phib(m-1), ..., \phib(0)\}$ via the definition of the effective action~\cite{itz} $A(\Phib,\Z(m))$
\be
A(\Phib,\Z(m)) = -C_m(\K) + \sum_{n=0}^m \k(n) \cdot \phib(n).
\ee
It is familiar that 
$
\frac{\partial A(\Phib,\Z(m))}{\partial \phi_a(n)} = k_a(n)
$
and the inverse of $\frac{\partial^2 C_m(\K)}{\partial k_a(n) \partial k_c(r)}$ is 
$
\gamma(\Phib,\Z(m))_{an,bn'} = \frac{\partial^2 A(\Phib,\Z(m))}{\partial \phi_a(n) \partial \phi_b(n')}.
$

Significant benefit comes from using the form of $A(\Phib)$ in
the formula for $C_m(\K)$ to find
\bea 
&&\exp[-A(\Phib,\Z(m))] = \\ \nonumber
&&\int \prod_{n=0}^m d^Dy(n) \exp[ \sum _{n=0}^m \k(n) \cdot (\y(n) - \phib(n))] \\ \nonumber
&&\exp[-A_0(\Y,\Z(m))], 
\eea
Considering the orbit $\phib(n)$ as a kind of base trajectory, change integration variables to 
$\psib(n) = \y(n) - \phib(n)$, denote 
$\Psib = \{\psib(m),\psib(m-1),...,\psib(0)\}$, and set $A_1(\Phib,\Z(m)) = A(\Phib, \Z(m)) - A_0(\Phib, \Z(m))$ leading to
 \bea 
&&\exp[-A_1(\Phib,\Z(m))] = \int \prod_{n=0}^m d^D\psi(n) \exp[ \sum _{n=0}^m \frac{A_1(\Phib,\Z(m)))}{\partial \phib(n)} \cdot \psib(n)] \nonumber \\
&& \exp[-\{A_0(\Phib + \Psib,\Z(m)) -A_0(\Phib,\Z(m)) \nonumber \\
&&-\sum _{n=0}^m \frac{A_0(\Phib,\Z(m))}{\partial \phib(n)} \cdot \psib(n) \}] 
\label{action}
\eea

If the $\Psib$ are small corrections to the base trajectory $\Phib$, we can expand the exponent in the integrand in powers 
of the $\psib(n)$, and to second order we have
\bea 
&&\exp[-A_1(\Phib,\Z(m))] = \nonumber \\
&&\int \prod_{n=0}^m d^D\psi(n) \exp[ -\frac{1}{2}\sum_{an,bn'}\psi_a(n) \gamma_0(\Phib,\Z(m))_{an,bn'}\psi_b(n')\nonumber \\
&&+ \sum_{n=0}^m\sum_{b=1}^D \frac{\partial A_1(\Phib,\Z(m))}{\partial \phi_b(n)} \psi_b(n)],
\eea
where $
\gamma_0(\Phib,\Z(m))_{an,bn'} = \frac{\partial^2 A_0(\Phib,\Z(m))}{\partial \phi_a(n)\,\partial \phi_b(n')}.
$
Performing the  Gaussian integral leads to the differential equation
\[
2A_1(\Phib,\Z(m)) =  \Delta 
- \frac{\partial A_1(\Phib,\Z(m))}{\partial \Phib} \gamma_0^{-1}(\Phib,\Z(m)) \frac{\partial A_1(\Phib,\Z(m))}{\partial \Phib},
\]
with $\Delta = \log\biggl [\frac{\det \gamma_0(\Phib,\Z(m))}{(2\pi)^{(m+1)D}}\biggr ]$.

We look for a solution to this differential equation as a power series in $\Phib$. The scalars $A_1(\Phib)$ and $\Delta(\Phib)$ (suppressing the dependence on $\Z(m)$) have the expansions $A_1(\Phib) = a_0 + a_1 \Phib 
+ \Phib a_2 \Phib/2 + \cdots$ and $\Delta(\Phib) = \Delta_0 + \Delta_1 \Phib + \Phib \Delta_2 \Phib/2 + \cdots$ to 2$^{nd}$ order 
in $\Phib$. 
$a_0, \Delta_0$ are scalars, $a_1, \Delta_1$ are vectors, and $a_2,\Delta_2$ are matrices. Each of these quantities depends on other aspects of the specific $A_0(\Phib, \Z(m))$, including $\Z(m)$. 

Comparing the terms in the differential equation results in $2a_0 = \Delta_0 - a_1  \gamma_0^{-1} a_1$, $2a_1 ({\cal I} + \gamma_0^{-1}a_2) = \Delta_1$, and $2a_2({\cal I} + \gamma_0^{-1}a_2) = \Delta_2$. ${\cal I}$ is the unit matrix. One may check that if $\Delta_1 = 0$ and $\Delta_2 = 0$, meaning that $A_0(\Phib)$ is quadratic, then consistently $a_1 = a_2 = 0$, and all probabilities are Gaussian. Higher order terms are determined by continuing the power series expansion.

More accurate approximations to Equation (\ref{action}) permit refinement of the differential equation for $A_1(\Phib,\Z(m))$.

Solving for $A_1(\Phib,\Z(m))$ allows us to address important questions such as the mean orbit during assimilation and the covariances about this mean. The mean trajectory through model phase space, call it $<y_a(n)> \,= v_a(n)$, is determined by 
\be
\frac{\partial A(\Phib,\Z(m))}{\partial \phi_a(n)}|_{v_a(n)} = 0.
\label{extremeorbit}
\ee
This stationarity of the effective action replaces the standard statement called 4D-VAR in the geosciences literature for determining the optimal orbit of the model system~\cite{lorenc}. 4DVAR uses $A_0(\Phib,\Z(m))$ alone.
 The covariance about $\v(n)$ is given by the inverse of the matrix $\gamma(\V,\Z(m))$ 
where $\V = \{\v(m),\v(m-1),...,\v(0)\}$. Higher moments are found through derivatives of $A(\Phib,\Z(m))$ evaluated along the orbit $\Phib = \V$. 

The mean phase space point reached at the end of the data assimilation window is $\v(m)$ and the $D \times D$ covariance about this mean location is $\gamma^{-1}(\V,\Z(m))_{am,bm}$. These, and possibly higher moments about $\v(m)$, are of physical interest, and any question of whether $P(\y(m)|\Z(m))$ is Gaussian~\cite{lorenc}, unlikely for any $A_0(\Y,\Z(m))$ of interest, is not under consideration.

If there are no model errors, then $P(\y(n+1)|\y(n)) = \delta^D(\y(n+1) - \f(\y(n),\p))$, and we can carry out all integrals in the expression for $C_m(\K)$ except that over $\y(0)$. This gives us
\bea
&&\exp[C^{NE}_m(\K(m))] = \int d^Dy(0) \exp[\sum_{n=0}^m \k(n) \cdot \y(n)] \nonumber \\
&&\exp[-A_0^{NE}(\Yhat,\Z(m))],
\eea
where $A_0^{NE}(\Yhat,\Z(m))$ is the same as $A_0(\Y,\Z(m))$ without the transition matrices and 
with $\y(n) = \f^{(n)}(\y(0),\p)$
and $\Yhat = \{\f^{(m)}(\y(0),\p), \f^{(m-1)}(\y(0),\p),...,\y(0)\}$.
This says that the trajectory starting from any of the points $\y(0)$ is carried to $\y(1) = \f(\y(0),\p)$, ..., 
and to $\y(m) = \f(\y(m-1),\p)$. All statistical information arises from $P(\y(0))$.

This is essentially the situation defined by~\cite{wett97} for differential equations and without assimilation of data. For discrete time dynamics, we have
\bea 
&&<y_a(n+1)> = \nonumber \\
&&<f_a(\y(n),\p)> = \biggl \{f_a(\frac{\partial }{\partial \k(n)},\p) \exp[C^{NE}_m(\K)]\biggr \}|_{\K = 0},
\eea
and this translates into an effective equation of motion for $\v(n)$: $\v(n+1) = \f^E(\v(n),\p)$,
where $\f^E(\phib,\p)$ is defined by a power series of derivatives on functions of $\phib$, then we use $\Phib = \V$. For example, if $\f(\y,\p)$ is quadratic in the state variables $\y$, so ($a,b,c = 1, ...,D$)
$
f_a(\y,\p) = A_a(\p) + B_{ab}(\p)y_b + C_{abc}(\p)y_b\,y_c,
$
then
$
f^E_a(\v(n),\p) = A_a(\p) + B_{ab}(\p)v_b(n) 
+ C_{abc}(\p)[v_b(n) v_c(n) + \gamma^{-1}(\V,\Z(m))_{bn,cn}]. 
$
A quadratic nonlinearity is likely to be interesting for many statistical fluid dynamics problems.

When there are model errors, we may incorporate them, to the extent they lead to reduced phase space resolution in the dynamics, by approximating the delta function transition probabilities as distributed Gaussians.

A summary of the path traversed here is that we followed routes established in statistical physics by transforming from a cumulant generating function for the mean orbit $<\y(n)>$ and covariances about that mean, to an effective action which lends itself to relative ease in its evaluation. We coupled these methods to the problem of estimating all state variables of a model in a setting where we have information from noisy measurements of a sparse subset of these, uncertainty about the model parameters, and uncertainty of the initial state when the measurements begin. 

For times later than the observation window $\{t_0,...,t_m\}$, all properties of the distribution of model variables $P(\y(t_M>t_m));\;M > m$ are determined by Equation (\ref{action}) with $TMI = 0$ for all times greater than $t_m$ where there are no new observations. We never have to explicitly evaluate the conditional probability distribution as the effective action provides us with the measurable expected values and covariances. We can add information from observations after $t_m$ through the addition of a nonzero conditional mutual information term whenever a measurement is provided using the same general formulation.

In the case of continuous labels in space and time for our dynamical variables the equation for $A_1(\Phib,\Z(m))$ would be functional. Indeed, attending to the fact that measurements are discrete in time and that numerical evaluations of  model state variables are also discrete in time, we have avoided many difficult mathematical questions. These have been addressed by Eyink and collaborators~\cite{eyink04}. 
The transition from our formulation of dynamics as a set of ordinary differential equations, discretized in time to iterated maps, to partial differential equations conveying information both in continuous space labels and in continuous time is not difficult in a formal sense. Indeed, statistical field theories often start from that perspective and simplify to our viewpoint. As measurements are actually performed in discretized time and coarse grained space, there may be advantage to beginning as we do.

The statistical physics framework we pose does suggest another route for exploration of its properties. One expects that as the sampling time and the spatial sampling lengths used in our discrete map dynamical equations become small, there will be scales of resolution below which there is not any improvement in the underlying physics. This suggests that a renormalization 
group~\cite{itz} formulation to express this independence of small scale phenomena may yield valuable approximations to the methods explored here.

Until this point, we have focused on the estimation of state variables by using probes $\K$ of the location of a model orbit as it incorporates information from measurements. One may estimate the fixed parameters in the model as well by asking that the effective action $A(\Phib)$ which is a function of those parameters also satisfy $\frac{\partial A(\Phib,\p)}{\partial \p} = 0$.
A formal argument can be given treating the fixed parameters as time dependent vectors $\p(n)$ satisfying $\p(n+1) = \p(n)$, but it illuminates little here. A more informative statement is that requiring the effective action to be stationary in the values of the fixed parameters is equivalent to asking that we maximize the total mutual information transferred from the measurements to the state of a model subject to imposition of the equations of motion $\y \to \f(\y,\p)$.

In our introduction we indicated that these methods may find use in the analysis of problems in a large number of different fields. The detailed association of equations of motion, representation of quantity and quality and statistics of measurements and their errors, and uncertainties in precision of initial states will vary with the specifics of the scientific arena being explored. 

Finally, the search over states and parameters to realize $\frac{\partial A(\Phib,\p)}{\partial \p} = 0$ and 
$\frac{\partial A(\Phib)}{\partial \phi_a(n)}|_{v_a(n)} = 0$,
when the dynamics leads to chaotic motions, requires regularization to permit numerical methods to be accurate~\cite{acg,ackr} and to allow estimation of the unobserved variables.

\section*{Acknowledgements}
This work was partially supported by the Office of Naval Research MURI Grant (ONR N00014-07-1-0741) and was completed during a visit to the Bernstein Center for Computational Neuroscience at the Ludwig Maxmillian's University (University of Munich), Munich, Germany.

\end{document}